\documentstyle[12pt,amssymb,psfig]{article}

\title{\bf A posteriori reading of Virtual Impactors 
impact probability}
\author{Germano D'Abramo\\
{\small Istituto di Astrofisica Spaziale e Fisica Cosmica,}\\
{\small Area di Ricerca CNR Tor Vergata, Roma, Italy}\\
{\small E--mail: {\tt dabramo@rm.iasf.cnr.it}}}
\date{}

\begin{document}

\maketitle

\begin{abstract} 

In this paper we define the {\em a posteriori} probability $W$. This
quantity is introduced with the aim of suggesting a reliable
interpretation of the actual threat posed by a newly discovered Near Earth
Asteroid (NEA), for which impacting orbital solutions (now commonly known
as Virtual Impactors or VIs) can not be excluded on the basis of the
available astrometric observations. The probability $W$ is strictly
related to the so-called background impact probability (that extrapolated
through a statistical analysis from close encounters of the known NEA
population) and to the annual frequency with which the impact monitoring
systems (currently NEODyS-CLOMON at University of Pisa and SENTRY at
NASA--JPL) find VIs among newly discovered asteroids. Of $W$ we also
provide a conservative estimate, which turns out to be of nearly the same
order of the background impact probability. This eventually says to us
what we already know: the fact that nowadays monitoring systems frequently
find VIs among newly discovered asteroids does not make NEAs more
threatening than they have ever been.

\end{abstract}

\section{Introduction}

Soon after a new Near Earth Asteroid (NEA) is discovered, a preliminary
orbit is computed using its positions in the sky over a suitable (minimal)
interval of time (astrometric observations). Like every physical
measurement, astrometric ones are affected by errors which make the
resulting orbit uncertain to some variable degree. Sophisticated
mathematical and numerical tools are currently available to orbit
computers which allow to propagate such measurement errors to the six
orbital elements which identify the orbit of the asteroid. For this
reason, the new NEA, soon after its discovery, is not represented by a
single point in the 6-dimensional dynamical elements space; rather, it is
represented by an uncertainty region, a 6-dimensional volume with diffused
contours. Obviously, the volume of this uncertainty region changes
(usually shrinks) when new more observations become available and the
orbit refines.

Moreover, when the nominal orbit of the new NEA is geometrically close to
the orbit of the Earth, and it shares some other peculiar orbital
characteristics, it can happen that some orbital solutions which lead to a
future collision of the asteroid with the Earth can not be excluded only
on the basis of the available astrometric observations. Namely, orbital
solutions which lead to a collision are inside the uncertainty region and
they are fully compatible with the available astrometric observations and
their errors.

What is substantially done in these cases by the researchers, with various
sophisticated techniques whose description is well beyond the scope of
this paper (see Milani {\it et al.}, 2000; 2003), is to sample the
uncertainty region according to a suitable frequency distribution (closely
related to what is currently known about error statistics) and then
evaluate the relative probability that the ``true'' orbit of the asteroid
is one of the collision ones. From now on we will refer to this
probability with the symbol ${\cal V}_i$. The collision orbits are
nowadays commonly called Virtual Impactors (or VIs; for an exhaustive
review see Milani {\it et al.}, 2003).

Every time new more astrometric observations become available, the quality
of the asteroid orbit improves and the estimated impact probability ${\cal
V}_i$ is re-computed. Its value is almost always such that ${\cal V}_i\ll
1$ and during the phases of the orbital refinement it fluctuates, usually
with a somewhat increasing trend\footnote{The reason of such increasing
behavior is rather technical and it is essentially connected to the fact
that uncertainty region shrinks with new more observations.} until it
falls to zero, its most probable final value.

Starting from 1999, some press announcements were made regarding as many
newly discovered NEAs which were found to have non zero collision chances
in the near future (given the highly chaotic dynamics involved in the
multiple planetary close encounters, which are at the basis of the impact
calculations, impact analysis procedures can safely cover only time spans
of the order of a century). The computations were carried out mainly by
two research groups, that at the University of Pisa and that at NASA--JPL.
One of the first and, maybe, most famous of such cases was that of
asteroid 1999~AN$_{10}$ (for more information, see for example Milani
{\it et al.}, 1999; 2003 and {\tt
http://impact.arc.nasa.gov/news/1999/apr/21.html}; for a detailed
historical account of these cases, see Chapman, 1999), being that of the
asteroid 2002~NT$_7$ the most recent one (up to this date). These objects
obviously rose the somewhat alarmed attention of the public opinion and of
the whole astronomical community for a while. Then, after the asteroids
orbits were refined thanks to new more astrometric observations and the
impact possibilities definitively ruled out, they have become again of
purely academic interest.

Currently, the only two existing automatized VIs monitoring systems,
CLOMON\footnote{{\tt http://newton.dm.unipi.it/neodys}} at University of
Pisa and SENTRY\footnote{{\tt http://neo.jpl.nasa.gov/risk/}} at
NASA--JPL, find tens of newly discovered NEAs with VIs orbital solutions
every year, and some with not so small impact probability (for a
preliminary statistics of VIs detections see Tab.~1, more later).

Given such past experience of public (and professional) reactions and
given the current rate of VIs orbital solutions discoveries, some
questions rise to the author's mind: how are VIs impact probabilities
actually related to the real impact threat? How much threat should we
reliably read in a VIs detection announcement? Equivalently, soon after
the discovery of VIs orbital solutions of a new asteroid, what is the
probability that ${\cal V}_i$ approaches and eventually reaches the unity
(within this paper we will use the compact notation ``${\cal V}_i \to
1$''), after the right amount of new more astrometric observations has
become available?

In this paper we give a statistical, {\em a posteriori} reading of VIs
impact probabilities (which actually are in their very nature
``deterministic'', or, more properly, {\em a priori})  in order to provide
an answer to such questions.

\section{Statistical reading of ${\cal V}_i$}

Soon after the discovery of VIs orbital solutions of a new asteroid, what
is the probability of ${\cal V}_i \to 1$, after the right amount of new
more astrometric observations has become available? It would seem quite
obvious that such probability is simply ${\cal V}_i$, according to the
definition of ${\cal V}_i$. But we believe that this is not the case. This
is essentially because ${\cal V}_i$, as we have said before, fluctuates
every time new more astrometric observations become available and the
computations are redone: which particular value should we consider for our
needs? The value obtained with the second batch of astrometric
observations following the discovery observations? Or the third? Or just
the value of ${\cal V}_i$ calculated with the discovery observations? And,
in latter case, if the discovery was made in another period of the year,
would the calculated value of ${\cal V}_i$ have been the same?  Actually,
what we are asking is: only knowing that a newly discovered NEA exhibits
some VIs orbital solutions, what is the probability that ${\cal V}_i$ will
be equal to $1$ at the end of the whole orbital refinement process? We
believe that only a statistical, {\em a posteriori} analysis can give an
acceptable and verifiable answer.

Now, let us make the following thought experiment, only functional to the
presentation of our point. Suppose that we are able to discover {\em all}
asteroids with absolute magnitude less than or equal to $H$ which pass
close to the Earth. Moreover, we reasonably suppose that every discovered
impacting asteroid will show some VIs, with low ${\cal V}_i$ soon after
the discovery and fluctuating with an increasing trend as soon as
subsequent astrometric observations become available. In other words, we
are putting ourselves in the somewhat idealized situation where {\em
every} impacting asteroid brighter than $H$ is surely discovered and for
it VIs monitoring systems surely spot some VIs soon after its discovery.

Thus, we define the {\em a posteriori} probability of ${\cal V}_i \to 1$,
which could be also interpreted as a kind of ``weight'' of the VIs impact
probability calculation (more on this later), as:

\begin{equation}
W(\leq H)= \frac{n(\leq H)}{v(\leq H)}\Biggr|_T=
\frac{\rho_{i}(\leq H)}{f_{{\cal V}_i} (\leq H)},
\label{eq1}
\end{equation}
where $n(\leq H)$ is the number of impacts of asteroids with absolute
magnitude less than or equal to $H$ and $v(\leq H)$ is the number of
asteroids with same absolute magnitude found among all the newly discovered
NEAs to exhibit VIs orbital solutions, both counted in the period of $T$
years. Note that, according to what we have said at the beginning of this
section, the number $n(\leq H)$ is counted in the number $v(\leq H)$,
since we have assumed that every impacting asteroid is identified soon
after its discovery as having some VI orbital solutions.

\begin{figure}[ht]
\centerline{\psfig{figure=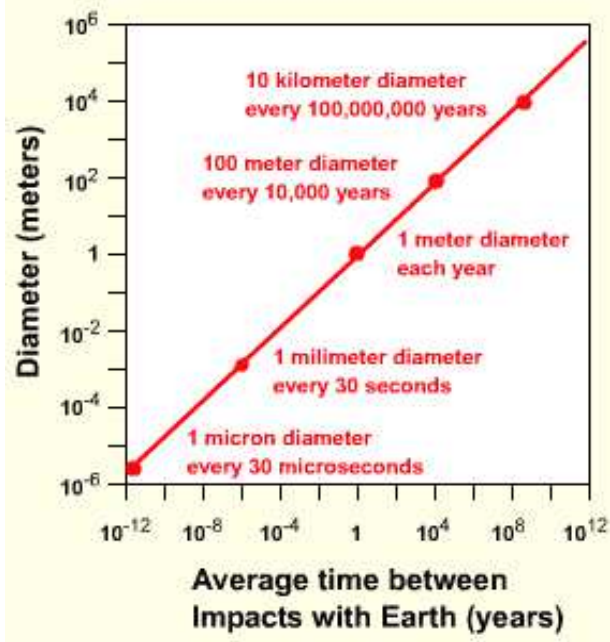,width=8cm,angle=0}}
\caption{An approximation of the average time between asteroidal impacts
with the Earth (the reciprocal of the background annual impact frequency
$\rho_{i}$)  as a function of the impactor's diameter. We choose to use
diameter, $D$, rather than absolute magnitude, $H$, since diameter is
a more direct physical quantity. The mathematical relation between $D$ and
$H$ is: $\log_{10}D\simeq \log_{10}1329 -
\frac{H}{5}-\frac{1}{2}\log_{10}p_V$, where $p_V$ is the albedo of the
asteroid and it is usually assumed to be equal to 0.15 (Chesley {\it et
al.}, 2002).}
\label{fig1} 
\end{figure}

Let us explain better the meaning of eq.~(\ref{eq1}). Within the
hypotheses introduced above on the almost perfect NEAs discovery
efficiency and VIs monitoring capabilities, we imagine to be able to wait
for a very long period of years, $T$, and count the number $n(\leq H)$ of
asteroid impacts and the number $v(\leq H)$ of VIs orbital solutions
detected among all the discovered NEAs below the characteristic absolute
magnitude $H$, within that period of time. Hence the fraction of these two
numbers gives the {\em a posteriori} probability that a newly discovered
asteroid, for which the VIs monitoring systems have spotted some VIs
orbital solutions, is just that which will fall on the Earth. In the third
member of eq.~(\ref{eq1}) we rewrite $W$ in terms of the background annual
impact frequency $\rho_{i}$, namely that extrapolated through a
statistical analysis from close encounters of the known NEA population
(Morrison {\it et al.}, 2003), and the annual frequency $f_{{\cal V}_i}$
of finding VIs among the newly discovered NEAs. A sketch of the average
time between impacts ($1/\rho_{i}$) is given in Fig.~\ref{fig1} as a
function of the impactor's diameter.

Before we have said that $W$ can be interpreted as a kind of ``weight'' of
the VIs impact probability calculation. This is because the greater is
$f_{{\cal V}_i}$, the lesser is the {\em a posteriori} probability of
${\cal V}_i \to 1$, no matter how is the initial numerical value of ${\cal
V}_i$. Namely, the higher is the frequency with which we find VIs among
newly discovered asteroids (with respect to the background frequency of
impacts, $\rho_{i}$), the lesser is their {\em a posteriori} weight in
expressing the threat of those particular newly discovered asteroids. This
mechanism shares inevitable analogies with the well-known ``crying wolf''
experience, bad faith apart.

Moreover, we can see that $W$ is not directly related to the specific
numerical value of ${\cal V}_i$. Rather, it depends upon $f_{{\cal V}_i}$,
namely the annual rate of VIs discovery, which, in turn, depends upon some
observational characteristics. These are the annual number of NEA
discoveries, the amount of astrometric observations available at
discovery, the magnitude of astrometric errors and the conventions in
their statistical treatment, as well as the observational geometry and
orbital characteristics of the newly discovered asteroid.  But, we guess,
it is not easy to give an exact estimate of its value at the moment.  A
larger sample of VIs detections is necessary in order to better estimate
$f_{{\cal V}_i}$ and thus the probability $W$.

Anyway, having an idea of the total number of VIs detections found at
every size between calendar years 2000 and 2001 (a sufficient but not
complete archive of VIs detections could be found in the Observing
Campaigns web pages of the Spaceguard Central Node at {\tt
http://spaceguard. rm.iasf.cnr.it}), a conservative estimate of $W$ cannot
be too much different from that we get with $f_{{\cal V}_i}$ of the order
of the unity. From Tab.~1 we see that the numerical value of $f_{{\cal
V}_i}$ varies between $\sim 1$ and $\sim 10$, in the reported range of
absolute magnitudes $H$. For the sake of simplicity, in this paper we
choose to adopt $f_{{\cal V}_i}\sim 1$ for all $H$, being confident of
committing a justifiable approximation, surely comparable with the
uncertainty with which $\rho_i$ is currently known (at least within some
ranges of absolute magnitude). With this reasonable assumption, we simply
have $W\sim\rho_{i}$.

Note that relaxing the optimistic assumptions on the almost perfect NEAs
discovery efficiency and VIs monitoring capabilities makes $f_{{\cal
V}_i}$, as approximated with the aid of Tab.~1, even an underestimate. And
consequently, it makes $W\sim\rho_{i}$ an overestimate.

Therefore, the probability of ${\cal V}_i\to 1$ is nearly of the order of
the background impact probability, no matter how is ${\cal V}_i$'s
specific, initial (fluctuating)  numerical value.

\begin{table}[t]
\begin{center}
\caption{Annual frequency of VIs detections below absolute magnitude $H$, 
estimated using the Spaceguard Central Node archive of VIs observing 
campaigns organized in the calendar years 2000 and 2001. During that period 
of time there were no VIs detections below $H=16$. In the reported range of
$H$, the numerical value of $f_{{\cal V}_i}$ varies between $\sim 1$ and 
$\sim 10$.} 
\vspace{1cm}
\begin{tabular}{l|c}
$H$ & $f_{{\cal V}_i}(\leq H)$   \\  
    & yr$^{-1}$  \\ \hline\hline
29 & 14.5 \\
28 & 14.0 \\
27 & 14.0 \\
26 & 13.5 \\
25 & 13.0 \\
24 & 9.5 \\
23 & 9.5 \\
22 & 8.5 \\
21 & 7.5 \\
20 & 7.0 \\
19 & 5.5 \\
18 & 3.5 \\
17 & 1.5 \\
16 & $\sim 1$\\
\hline\hline
\end{tabular}
\end{center}
\end{table}

This result should not be a surprise since it simply re-states what we
already know: the actual impact threat of a newly discovered NEA
exhibiting VIs orbital solutions is always the same estimated through the
close encounters statistics of known population. The fact that in the last
few years many VIs orbital solutions have been detected among newly
discovered NEAs obviously does not make NEAs more threatening than they
have ever been.

As a matter of fact, the annual rate of VIs detections, if compared with
the background impact probability, suggests an order of magnitude of the
weight VIs detections have in expressing the real threat of a newly
discovered NEA with ${\cal V}_i\neq 0$.

\section{Conclusions}

From the definition and the discussion of the {\em a posteriori}
probability $W$ done in this paper it follows that, rigorously speaking,
the VIs impact probability ${\cal V}_i$ does not give the real expression
of the actual impact threat posed by a newly discovered NEA exhibiting VIs
orbital solutions. This is properly done by $W$, which is strictly related
to the so-called background impact probability $\rho_{i}$ (that
extrapolated through a statistical analysis from close encounters of the
known NEA population) and to the annual frequency with which the impact
monitoring systems (currently NEODyS-CLOMON at University of Pisa and
SENTRY at NASA--JPL) find VIs among newly discovered asteroids. Of $W$ we
also provide a conservative estimate, which turns out to be of nearly the
same order of $\rho_{i}$.

All this might seem a bit paradoxical, given the definition of ${\cal
V}_i$. Yet, a closer look at the definition of $W$ shows that our
conclusions are straightforward and even obvious.

%\section*{Acknowledgments}

\section*{References}

\noindent Chapman, C.R., 1999. 
The asteroid/comet impact hazard. Case Study for 
{\it Workshop on Prediction in the Earth Sciences: Use and Misuse in Policy 
Making}, July 10-12 1997 - Natl.~Center for Atmospheric Research, Boulder, CO
and September 10-12 1998, Estes Park, CO. Available on-line at:
{\tt http://www.boulder.swri.edu/clark/ncar799.html}\\

\noindent Chesley, R.S., Chodas, P.W., Milani, A., Valsecchi, G.B.,
Yeomans, D.K., 2002. Quantifying the risk posed by potential Earth
impacts. {\it Icarus}, in press.\\

\noindent Milani, A., Chesley, S.R., Valsecchi, G.B. 1999. Close 
approaches of asteroid 1999 AN$_{10}$: resonant and non-resonant returns.
{\it Astronomy \& Astrophysics} 346:L65-L68.\\

\noindent Milani, A., Chesley, S.R., Valsecchi, G.B. 2000. Asteroid close 
encounters with Earth: Risk assessment. {\it Planetary \& Space Science} 
48: 945-954.\\ 

\noindent Milani, A., Chesley, S.R., Chodas, P.W., Valsecchi, G.B.
Asteroid close approaches and impact opportunities. Chapter for {\it
Asteroids III} book edited by William Bottke, Alberto Cellino, Paolo
Paolicchi, and Richard P. Binzel. University of Arizona Press, Tucson
(2003).\\

\noindent Morrison, D., Harris, A.W., Sommer, G., Chapman, C.R., Carusi,
A. Dealing with the Impact Hazard. Chapter for {\it Asteroids III} book
edited by William Bottke, Alberto Cellino, Paolo Paolicchi, and Richard P.
Binzel. University of Arizona Press, Tucson (2003).

\end{document}